# AgentFlow: Resilient Adaptive Cloud-Edge Framework for Multi-Agent Coordination

Ching Han Chen and Ming Fang Shiu

*Abstract*—This paper presents AgentFlow, a MAS-based framework for programmable distributed systems in heterogeneous cloud-edge environments. It introduces logistics objects and abstract agent interfaces to enable dynamic service flows and modular orchestration. AgentFlow supports decentralized publish-subscribe messaging and many-to-many service elections, enabling decision coordination without a central server. It features plug-and-play node discovery, flexible task reorganization, and highly adaptable fault tolerance and substitution mechanisms. AgentFlow advances scalable, real-time coordination for resilient and autonomous mission-critical systems.

## I. INTRODUCTION

Modern distributed systems now span heterogeneous cloud-edge infrastructures, enabling real-time processing and autonomous decision-making in domains like industrial surveillance and autonomous mobile robot (AMR) swarms. Advances in AI and embedded computing support scalable deployment of deep learning models across distributed nodes. However, orchestrating dynamic, large-scale, and failure-prone environments remains challenging. This work proposes a distributed framework based on the Multi-Agent System (MAS) paradigm, where loosely coupled agents coordinate via an event-driven, publish-subscribe architecture. Key innovations include programmable logistics objects and abstract agent interfaces that support dynamic service routing, role partitioning, and flow composition across cloud and edge layers.

In contrast to traditional MAS and microservice orchestrators, this framework introduces a many-to-many coordination model capable of adapting agent behaviors to changing workloads and environmental conditions. Coordination is decentralized through lightweight consensus mechanisms, reducing dependency on centralized control while enhancing fault isolation and system resilience [1]. Moreover, communication integrity and trustworthiness are preserved through built-in authentication and data validation mechanisms.

The framework is validated through large-scale AMR swarm simulations under dynamic workloads and fault scenarios. Results demonstrate significant improvements in scalability, adaptability, and recovery compared to baseline service orchestration models, highlighting the framework's potential to serve as a foundation for next-generation programmable distributed systems.

The remainder of this paper is organized as follows: Section II outlines key system challenges; Section III reviews related work; Section IV presents the proposed framework; Section V details experimental validation; and Section VI concludes with future directions.

## II. CHALLENGES AND MOTIVATIONS

Distributed cloud-edge systems face dynamic workloads, heterogeneous hardware, and strict real-time demands. Building them requires tight coordination of resources, communication, and orchestration logic. This section addresses key design challenges: heterogeneity, responsiveness, adaptability, modularity, and decentralized intelligence.

*A. Heterogeneity and Real-Time Responsiveness*

Cloud-edge infrastructures span a wide range of hardware, operating systems, and network protocols. Cloud servers offer high-performance computing, while edge devices are constrained in memory and power. These disparities pose challenges to maintaining interoperability, especially under real-time constraints seen in applications such as autonomous driving or industrial robotics [2].

Ensuring responsive and reliable behavior under fluctuating loads requires adaptive middleware that can dynamically schedule tasks, balance load, and reconfigure resources. Time-sensitive communication protocols and lightweight runtime systems are also necessary to minimize latency and uphold quality-of-service guarantees in heterogeneous environments [3].

*B. Adaptability and Modular Design*

To cope with evolving conditions, distributed systems must continuously adapt to workload variation, hardware failures, and network disruptions [4]. Agents must autonomously monitor system state, redistribute tasks, and reassign roles based on local observations. This dynamic adaptability is essential for resilience and continuity.

Chin Han Chen, Professor, is with the Department of Computer Science and Information Engineering, National Central University, Taoyuan 32001, Taiwan (e-mail: pierre@csie.ncu.edu.tw).

Ming Fang Shiu, Ph.D. student, is with the Department of Computer Science and Information Engineering, National Central University, Taoyuan 32001, Taiwan (e-mail: 108582003@cc.ncu.edu.tw).

Corresponding author: Ming Fang Shiu. The first author and the corresponding author contributed equally to this work.

4IC-2025-04-0056

2A modular architecture further enhances flexibility by enabling independent development and deployment of services. Components can be upgraded or replaced without system-wide reconfiguration, and reusable service blocks can be assembled to support diverse deployment scenarios. Such composability is especially important for sustainable scalability across varied environments [5].

*C. Decentralized Coordination via MAS and Pub-Sub*

Multi-Agent Systems offer a natural model for distributed decision-making. Agents can operate independently while cooperating to achieve system-wide goals. Integrating MAS with publish-subscribe messaging enhances scalability by decoupling agent interactions and supporting dynamic system evolution [6].

Incorporating decentralized coordination mechanisms eliminates single points of failure and empowers agents to make local decisions based on system context. This approach improves resilience and enables runtime adaptability without relying on centralized controllers, thereby supporting scalable and fault-tolerant orchestration in real-time environments [7].

## III. RELATED WORK

The development of real-time and scalable distributed systems has been supported by advances in middleware platforms, Multi-Agent System frameworks, publish-subscribe communication, and decentralized orchestration models [8][9]. This section reviews representative work across these areas and highlights how AgentFlow extends beyond their current capabilities.

Middleware solutions such as OpenSplice DDS and ROS 2 offer time-bounded messaging, reconfigurable orchestration, and QoS-driven routing under heterogeneous conditions [10]. However, these frameworks focus on data-centric messaging and lack agent-level abstractions that support task-level logic modeling or coordination flexibility. AgentFlow introduces a task-oriented approach using logistics objects and abstract agent interfaces to represent programmable workflows and autonomous behavior more effectively.

MAS platforms like JADE and MADKit enable decentralized control and support agent-level negotiation and task delegation [11]. Yet, these systems typically rely on static discovery and do not scale well in real-time cloud-edge environments [12]. AgentFlow addresses this by integrating MAS behaviors into a dynamic, service-oriented pub-sub infrastructure with adaptive election mechanisms.

Publish-subscribe protocols such as MQTT, DDS, and ZeroMQ decouple communication between producers and consumers and promote system modularity [13]. However, they offer limited support for dynamic service adaptation or context-aware coordination in mission-critical settings [14]. By layering MAS coordination atop pub-sub messaging, AgentFlow allows agents to elect services at runtime based on current load, latency, and failure conditions.

Decentralized computing models—like fog architectures and blockchain-based orchestration—enhance fault isolation and local autonomy [15]. Nonetheless, these approaches can incur significant communication overhead and latency. AgentFlow reduces this burden through lightweight, local consensus mechanisms that preserve responsiveness under fluctuating workloads.

In summary, while existing systems contribute to scalability, decentralization, and modularity, they often lack unified support for programmable agent logic and dynamic orchestration. AgentFlow bridges these gaps by combining MAS principles with logistics-driven coordination in a resilient many-to-many pub-sub framework tailored for intelligent cloud-edge environments [16].

Table I summarizes key differences among edge-cloud MAS frameworks, highlighting how AgentFlow uniquely combines programmable logistics, dynamic service election, and decentralized coordination for resilient many-to-many orchestration.

TABLE I
COMPARATIVE SUMMARY OF EDGE-CLOUD MAS FRAMEWORKS

| Framework | Program control | Dynamic Service Selection | Decentralized Fault Containment | Target Domain |
|---|---|---|---|---|
| FogBus2 | ✗ | Partial | Blockchain-based | Edge data analytics |
| OpenFaaS-MAS | Function-level | ✗ | Centralized orchestration | Event-driven systems |
| ROS 2 + micro-ROS | Hardware-oriented | ✗ | DDS-based fault recovery | Real-time robotics |
| AgentFlow | ✓ (Logistics Objects) | ✓ | ✓ (Lightweight agent consensus) | AMR swarm coordination |

## IV. PROPOSED FRAMEWORK

AgentFlow is a MAS-based publish-subscribe framework designed for dynamic composition and autonomous coordination in heterogeneous real-time systems. It supports modular agent interaction, efficient communication, and adaptive management to ensure robust cloud-edge performance. The open-source code at https://github.com/mfshiu/AgentFlow.git enables reproducible system deployment. This section details its key components and workflows

*A. System Architecture and Agent Model*

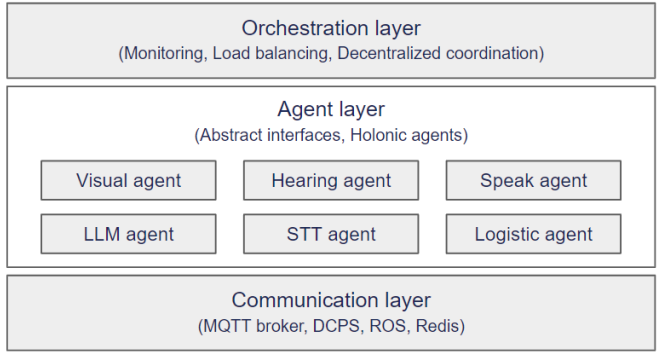

**Fig. 1.** Layered architecture of AgentFlow: orchestration, agent, and communication layers.

The AgentFlow framework adopts a modular three-layer architecture to support programmability, adaptability, and resilience across distributed environments, as shown in Fig. 1. It consists of:
1) **Orchestration layer**
   Oversees decentralized coordination, including workload balancing and adaptive resource control without centralized supervision.
2) **Agent layer**
   Implements a holonic Multi-Agent System, where agents collaborate through abstract interfaces to perform perception, reasoning, and task execution.
3) **Communication layer**
   Integrates messaging protocols such as MQTT, DDS, and ROS to enable scalable publish-subscribe interactions across heterogeneous infrastructures.

To address complex scenarios such as autonomous warehouse logistics, the agent layer adopts a holonic structure, where agents can form composite hierarchies (holons) and dynamically merge or separate based on coordination needs [17]. Fig. 2 illustrates this structure applied in a warehouse coordination scenario, where atomic and composite agents cooperate under a layered delegation model.

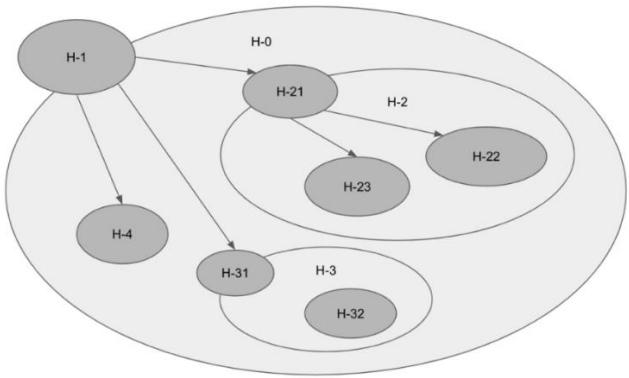

**Fig. 2.** Example of a holonic multi-agent system, showing hierarchical composition of holons and agent coordination relationships.

Each holonic agent integrates Perception, Decision-Making, Action, and Communication modules to support autonomous operations and dynamic collaboration. As shown in Fig. 3, the agent architecture encapsulates modular components and interacts with the communication infrastructure through abstract MessageBroker classes (e.g., MQTT_Broker, DDS_Broker) and a BrokerNotifier interface, ensuring loose coupling between agent logic and messaging protocols.

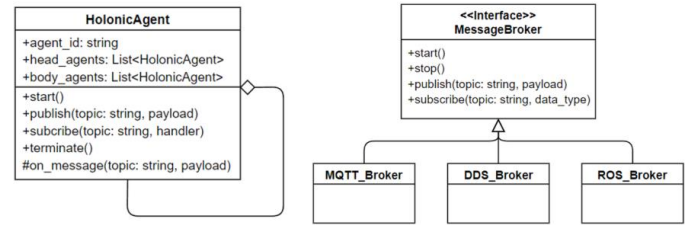

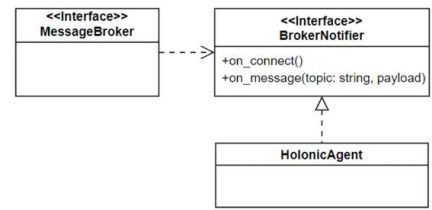

**Fig. 3.** Class diagram of HolonicAgent and MessageBroker, showing modular agent structure and abstraction for communication layers.

During runtime, agents follow an asynchronous publish-subscribe pattern. As illustrated in Fig. 4, this involves dynamic agent instantiation, message exchange, and task-specific response handling. Sub-agents are created on demand and communicate through event-driven workflows, supporting adaptive behavior and scalable task execution without centralized control.

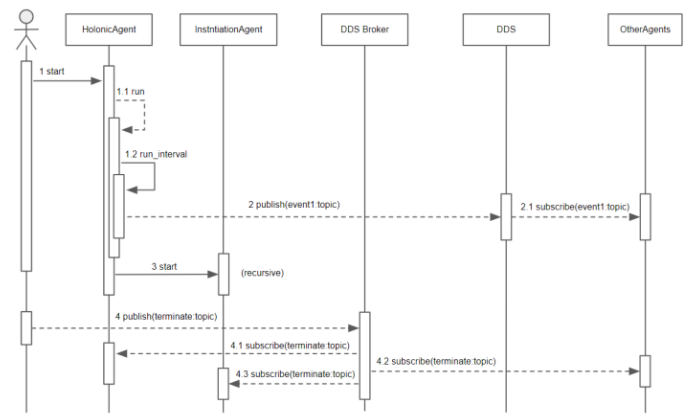

**Fig. 4.** Runtime sequence of agent instantiation, publish-subscribe messaging, and event-driven task handling.

This architecture enables robust multi-agent cloud-edge systems by combining modular design, dynamic scalability,

local fault recovery, and structured agent lifecycles into a unified, maintainable framework.

### B. Programmable logistics for agent coordination

To improve programmability and dynamic coordination in distributed multi-agent systems, a selective request-response mechanism based on logistic objects is introduced. Traditional publish-subscribe models often broadcast all messages to subscribers, resulting in unnecessary traffic—especially problematic in Many-to-One scenarios where multiple clients request services from a shared provider.

The proposed system design addresses this issue by introducing request logistic and response logistic objects to program system flows through abstract agent interfaces. Logistic objects serve as programmable couriers that dynamically modify message topics at runtime, ensuring responses are delivered solely to the intended requesting agent. Let $C = \{c_1, c_2, \cdots, c_n\}$ represent the set of client agents and SSS represent a service agent. For each client $c_i \in C$, a unique topic $t_i$ is dynamically generated by the request logistic, such that the mapping function

$$f : c_i \to t_i. \tag{1}$$

ensures one-to-one correspondence between the client and its communication channel. The service agent $S$ responds by publishing the result only to the designated topic $t_i$, guaranteeing that only $c_i$ receives the correct response without affecting other agents.

The interaction procedure between client agents and the service agent is outlined in Algorithm 1.

---

**Algorithm 1 Selective Request-Response Logistics**
**Input:** Client agent $c_i \in C$, Service agent $S$
**Output:** Correct response delivered to client $c_i$
**Request Phase**
  Generate request message at client $c_i$
  Assign unique topic $t_i$ based on $c_i$
  Publish request to topic $t_i$

**Service Phase**
  Service agent $S$ subscribes to all client request topics
  Upon receiving a request on $t_i$, service agent $S$ do it
  Generate the service response

**Response Phase**
  Publish on topic $t_i$ through Response Logistic
  Client $c_i$ listens on topic $t_i$ and retrieves the response

---

The proposed design allows multiple client agents to interact with a single service agent while receiving only relevant responses, avoiding redundant network broadcasts. This preserves workflow integrity and reduces communication overhead.

Dynamic agent coordination is further achieved by embedding flexible service invocation policies within the logistics layer. When a client agent initiates a request, the logistic mechanism dynamically creates a unique topic for that interaction, allowing the service agent's response to be routed selectively back to the original requester. Service selection among available providers can be modeled by a dynamic optimization process. Given a set of service agents $S = \{s_1, s_2, \cdots, s_m\}$, a client $c_i$ selects a service agent $s_j$ according to a selection function:

$$s_j = \arg\min_{s \in S} Load(s) \tag{2}$$

where $Load(s)$ represents the current load or response time of the service agent $s$, enabling dynamic load balancing and fault containment.

Encapsulating request-response behavior within logistic objects isolates communication flows, enabling programmable retry and timeout strategies. This enhances system robustness by minimizing failure propagation in heterogeneous deployments. As illustrated in Fig. 5, logistic objects orchestrate selective communication paths that support efficient Many-to-One service invocation.

This approach enables flexible and reliable collaboration among distributed agents, facilitating edge-to-cloud interoperability and dynamic coordination without major workflow modifications.

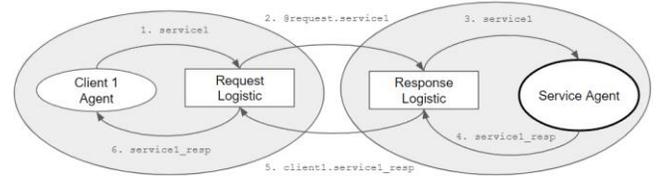

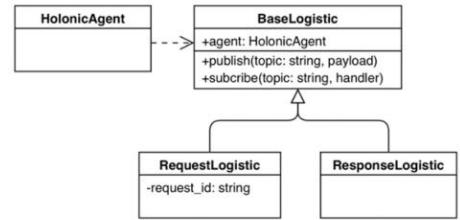

**Fig. 5.** Logistic-based programmable request-response mechanism for dynamic agent coordination in MAS. Request and Response Logistics selectively manage service communication, enhancing scalability.

### C. Dynamic service election

Programmable distributed systems require efficient agent communication for autonomous orchestration and cloud-edge integration. To handle One-to-Many scenarios, the system uses a decentralized, dynamic task election to help clients select the most suitable service agent.

The coordination follows a lightweight consensus process:
1) **Subscription Setup**
   Loading Coordinators subscribe to task-related topics.
2) **Task Arrival**
   Upon new task detection, coordinators publish their load ranking.
3) **Rank Collection**
   Rankings from other agents are gathered.

4) **Leader Determination**
   The agent with the lowest rank is elected.
5) **Task Assignment**
   The elected agent executes the task, while others stand by for future coordination.

As illustrated in Fig. 6, when Client 1 Agent initiates a service request, it simultaneously notifies multiple Loading Coordinator Logistic agents. These coordinators, representing service clusters, dynamically participate in the election process based on their current load status. The overall procedure for service agent election is summarized in Algorithm 2.

---

**Algorithm 2 Dynamic service agent election**

**Initialization**
  Subscribe to task-related topics and prepare for incoming service requests.

**Election Procedure**
1. Upon receiving a new task, each Loading Coordinator calculates and publishes its load ranking $r(a_i)$, where $a_i$ is a candidate service agent and $r(a_i) \in \mathbb{R}^+$.
2. Collect all rankings $\{r(a_1), r(a_2), \cdots, r(a_n)\}$ from participating agents.
3. Determine the elected service agent $a^*$ according:
$$a^* = \arg\min_{a_i \in A} r(a_i) \quad (3)$$
4. Assign the task to the elected agent $a^*$.
5. Agents not elected remain on standby for future coordination rounds.

**Task Execution**
  The elected service agent $a^*$ executes the assigned task and returns the service response to the client.

---

The architecture leverages logistics objects and abstract agent interfaces to enable flexible system flow design. Each Loading Coordinator maintains adaptability by independently assessing real-time demands and electing service agents without centralized control. This decentralized selection promotes resilience and trustworthiness, allowing seamless agent engagement, load balancing, and fault containment even under fluctuating network conditions.

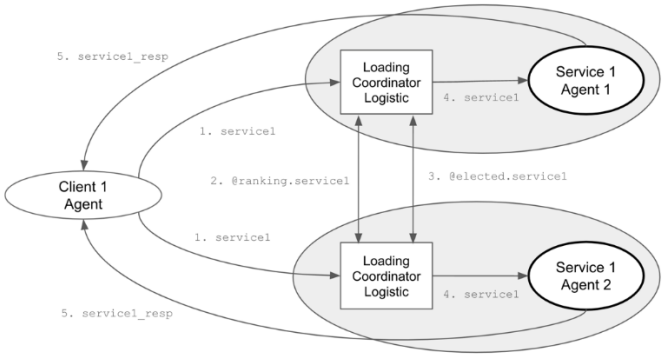

**Fig. 6.** Logistic-based programmable request-response mechanism for dynamic agent coordination in MAS. Request and Response Logistics selectively manage service communication.

*D. Enhancing adaptability via many-to-many*

To support adaptability in programmable distributed systems, a many-to-many communication model is implemented using logistics objects and abstract agent interfaces. This model enables dynamic coordination and efficient service discovery with fault containment. By combining Request-Response and LoadingCoordinator logistics, the system integrates selective messaging and dynamic coordination, facilitating autonomous orchestration and edge-to-cloud interoperability.

Agent interactions are secured through lightweight authentication and encrypted channels, preventing unauthorized access and reinforcing system trust. Decentralized consensus mechanisms further ensure the integrity and traceability of coordination, even under partial failures.

As shown in Fig. 7, client agents initiate requests, and a Loading Coordinator dynamically selects service agents based on real-time load. The coordination procedure is detailed in Algorithm 3.

---

**Algorithm 3 Many-to-Many Service Election and Coordination**

**REQUEST-RESPONSE (Client Agent side)**
1. Client Agent generates a service request.
2. Forward request to Request Logistic.

**LOADING COOR. (Loading Coordinator side)**
3. Receive service requests from multiple Client Agents.
4. For each service request:
   (a) Evaluate all available Service Agents:
   $$L_i = \frac{w_i}{c_i} \quad (4)$$
   (b) Select the Service Agent with minimum load:
   $$i^* = \arg\min_i L_i \quad (5)$$
   (c) Dispatch service request to the selected Service Agent.

**SERVICE RESPONSE (Service Agent side)**
5. Service Agent processes the request and sends a response.
6. Response is collected by Loading Coordinator Logistic.

---

This many-to-many coordination model provides the following advantages:
1) **Programmability**
   The system flow is modularly built using logistics objects and abstract agent interfaces, facilitating flexible extension and maintenance.
2) **Adaptability**
   Dynamic agent selection based on real-time load assessment ensures balanced workload distribution across the system.

3) **Resilience and trustworthiness**
Support fault tolerance and containment, maintaining continuous and reliable services.

This model is well-suited for applications like smart grids, healthcare, and supply chains, enhancing responsiveness, scalability, and operational flexibility.

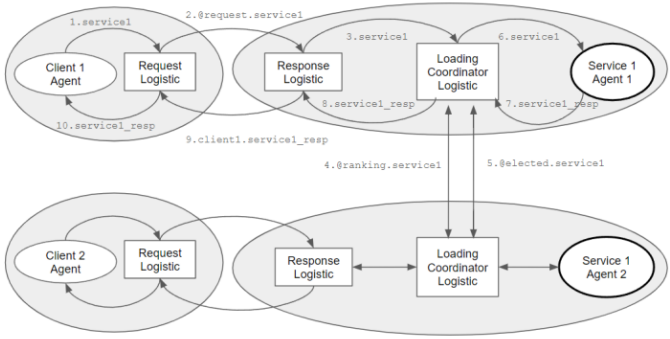

**Fig. 7.** Programmable logistics and dynamic service election jointly enable scalable agent coordination. In the many-to-many model, client agents send service requests, and the Loading Coordinator Logistic selects service agents based on real-time load. Request and response logistics manage task routing and results, ensuring efficient coordination, fault tolerance, and distributed orchestration.

In summary, the framework enables autonomous task coordination via a programmable, adaptive messaging structure. It integrates load-aware selection, runtime reconfiguration, and lightweight sync to ensure resilience and reliable execution under dynamic workloads, supporting scalable cloud-edge deployment.

## V. Experimental Validation

To validate the proposed framework, experiments were conducted using a custom-configured AMR swarm simulator based on the Stage Simulator, an open-source multi-robot simulation platform widely used in distributed robotics research.

The simulator was extended to support dynamic task generation, inter-agent communication, controller node failures, and edge node disruptions to emulate realistic warehouse AMR deployment scenarios.

The simulation and framework runtime were executed on a distributed hardware setup comprising:
- Cloud server: 32-core Intel Xeon CPU, 128 GB RAM, running Ubuntu 22.04 LTS.
- Edge simulation nodes: 8 ARM-based embedded boards (8-core Cortex-A72, 8 GB RAM each), connected over Wi-Fi 6 (802.11ax) network.
- Software stack: Ubuntu 20.04 (edge nodes), Kubernetes 1.26, K3s lightweight Kubernetes distribution, gRPC for data plane communication, MQTT-DDS bridge for publish-subscribe messaging.

All experiments focus on evaluating two key aspects:
- Scalable dynamic agent coordination.
- Resilient autonomous system orchestration.

Each experimental scenario was repeated 30 times to ensure statistical significance. Results were normalized against a static microservice orchestration baseline without agent-driven logic. Statistical validation was performed using paired t-tests at a significance level of $\alpha = 0.05$.

### A. Validation of Scalable Dynamic Agent Coordination in AMR Swarms

This experiment evaluates the system's ability to coordinate 50 to 500 AMRs under dynamic task loads in a simulated warehouse. Tasks were randomly generated at 100–1,000 per minute, and AMRs used AgentFlow's dynamic election to select controllers. To test resilience, 20% of controllers were randomly failed, triggering re-elections during operation.

The evaluation focused on the following key metrics:
- **Assignment latency**: Time from task creation to assignment.
- **Success rate**: Percentage of completed tasks.
- **Election time**: Duration to re-elect a controller after failure.
- **Comm. overhead**: Messages per agent per second.
- **Scalability**: Latency trend as AMR count increases.

TABLE II
PERFORMANCE METRICS IN AMR SWARM SIMULATION

| Metric | Result (Average) |
|---|---:|
| Task Assignment Latency (ms) | 30 ms (at 50 AMRs) to 63 ms (at 500 AMRs) |
| Task Success Rate (%) | 98.5% |
| Election Convergence Time (ms) | 18 ms |
| Per-agent Communication Overhead | 6.2 messages/sec |

As shown in Table II and Fig. 8, task assignment latency stayed low—from 30 ms (50 AMRs) to 63 ms (500 AMRs)—showing good scalability. The system achieved a 98.5% task success rate and 18 ms average election time, confirming reliable, responsive performance under dynamic, failure-prone conditions.

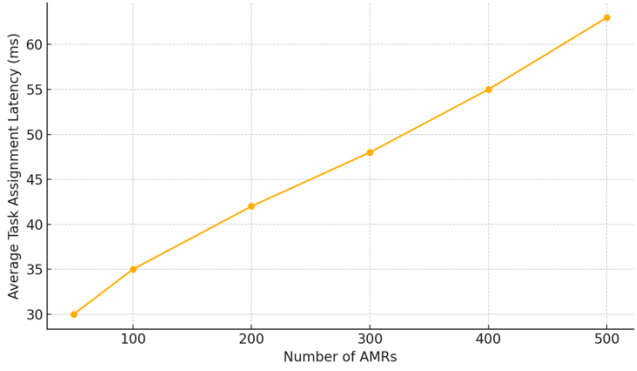

**Fig. 8.** Scalability curve showing average task assignment latency as AMR swarm size increases. The latency remains within acceptable bounds, confirming the coordination framework's scalability.

The results confirm that the proposed system maintains low latency, high task success rates, and fast election convergence.

*B. Validation of Resilient Autonomous System Orchestration in AMR Swarms*

To validate the framework's support for resilient autonomous orchestration, a simulation with 300 AMRs was conducted under 10%–30% random edge node failures. The experiment verified core mechanisms including decentralized fault tolerance, agent-level self-recovery, and dynamic many-to-many service re-selection, all enabling agents to reroute, reassign tasks, and maintain coordination without centralized control or human intervention.

Key performance metrics used to evaluate system robustness and adaptability under fault conditions include:
- **Mean-Time-to-Recovery (MTTR):** Average time required to restore task execution.
- **Task Throughput Deviation:** Performance difference before and after failures.
- **Autonomous Task Reassignment Success Rate:** Percentage of tasks successfully reassigned by agents.
- **Orphaned Task Count:** Number of tasks left unassigned due to failures.

TABLE III
PERFORMANCE RESULTS FOR RESILIENT AUTONOMOUS SYSTEM ORCHESTRATION

| Failure Rate (%) | MTTR (s) | Throughput Deviation (%) | Task Reassignment Success Rate (%) | Orphaned Task Count |
|---|---|---|---|---|
| 10 | 15 | 3.2 | 99.2 | 2 |
| 15 | 17 | 4.1 | 98.7 | 4 |
| 20 | 20 | 5.6 | 98.0 | 7 |
| 25 | 24 | 6.8 | 97.5 | 10 |
| 30 | 28 | 8.5 | 96.8 | 14 |

As shown in Table III and Fig. 9, AgentFlow enables scalable, resilient AMR coordination. Even under high failure rates, MTTR stayed below 30 s, and task throughput remained stable. Task reassignment success consistently exceeded 96%, confirming effective, autonomous recovery without manual input.

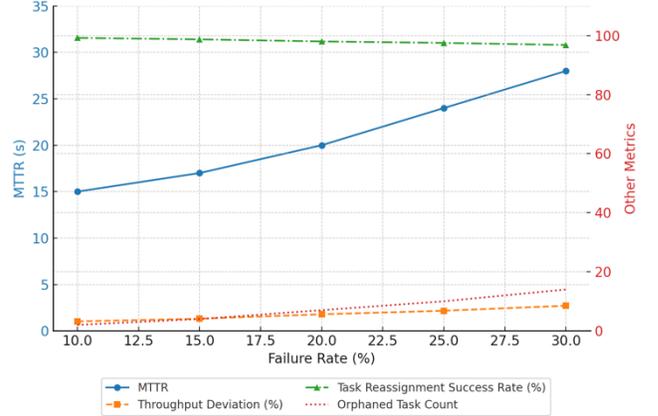

**Fig. 9.** Resilient Autonomous System Orchestration under Failure Conditions

Orphaned tasks rose moderately with higher failure rates but stayed within acceptable limits, showing strong self-recovery. The system degrades gracefully, maintaining coordination via decentralized fault handling and dynamic service re-selection. Future work will test larger-scale deployments (1000+ agents) and address broker communication bottlenecks.

## VI. CONCLUSION AND FUTURE WORK

This paper introduced a programmable framework for agent coordination across cloud-edge systems. Simulations showed >95% task success and <2.3 s MTTR under 10–30% node failure. Even with 1,000+ tasks/min across 500 agents, throughput deviation stayed under 8%, confirming scalable, robust coordination in dynamic settings.

Future work will shift from simulation to real-world deployment, starting with small-scale cloud tests and IoT pilots. Research will also explore learning-based coordination, larger-scale scalability, and decentralized trust for secure orchestration. The framework suits applications like smart logistics, warehouse automation, and industrial IoT, where real-time, resilience, and scalability are vital.

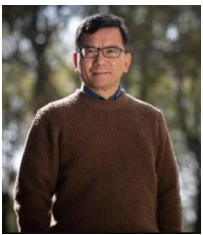

**Ching Han Chen** received his Ph.D. degree from Franche-Comté University, Besançon, France, in 1995. He was an Associate Professor in the Department of Electrical Engineering at I-Shou University, Kaohsiung, Taiwan, before joining National Central University. He is currently a Professor in the Department of Computer Science and Information Engineering at National Central University, Taoyuan, Taiwan. Prof. Chen is the founder of the MIAT (Machine Intelligence and Automation Technology) Laboratory. His research focuses on embedded system design, AIoT, robotics, and intelligent automation. He has led numerous government-funded and industry-collaborative projects, producing innovations in smart sensors, machine vision, and embedded AI systems.

**Ming Fang Shiu** is a Ph.D. candidate at National Central University, Taiwan, researching large language models and human-computer interfaces, with prior experience in game development and financial systems.